\documentclass[aps,prl,reprint,superscriptaddress,showpacs]{revtex4-1}

\usepackage{color,graphicx,amssymb,amsmath}

\newcommand{\veps}{\varepsilon}
\newcommand{\mr}[1]{\mathrm{#1}}
\newcommand{\mb}[1]{\mathbf{#1}}
\renewcommand{\k}{\mathbf{k}}

\newcommand{\el}{|e|}

\begin{document}


\title{Anisotropic ultrafast electron dynamics induced by high-field terahertz pulses in $n$-doped InGaAs}



\author{F. Blanchard}
\affiliation{INRS-EMT, Advanced Laser Light Source, Universit\'e du Qu\'ebec, Varennes, Qu\'ebec J3X 1S2, Canada}

\author{D. Golde}
\affiliation{Department of Physics and Materials Sciences Center, Philipps-University, Renthof 5, 35032 Marburg, Germany}

\author{F. H. Su}
\affiliation{Department of Physics, University of Alberta, Edmonton, Alberta T6G 2G7, Canada}

\author{L. Razzari}
\author{G. Sharma}
\author{R. Morandotti}
\author{T. Ozaki}
\affiliation{INRS-EMT, Advanced Laser Light Source, Universit\'e du Qu\'ebec, Varennes, Qu\'ebec J3X 1S2, Canada}

\author{M. Reid}
\affiliation{Department of Physics, University of Northern British Columbia, Prince George, British Columbia V2N 4Z9, Canada}

\author{M. Kira}
\author{S. W. Koch}
\affiliation{Department of Physics and Materials Sciences Center, Philipps-University, Renthof 5, 35032 Marburg, Germany}

\author{F. A. Hegmann}
\email[]{hegmann@ualberta.ca}
\affiliation{Department of Physics, University of Alberta, Edmonton, Alberta T6G 2G7, Canada}



\date{\today}

\begin{abstract}
The anisotropic effective mass of electrons is directly measured using time-resolved THz-pump/THz-probe techniques in a $n$-doped InGaAs semiconductor thin film. 
A microscopic theory is used to attribute this anisotropy in the THz probe transmission to the nonparabolicity of the conduction band. 
Self-consistent light-matter coupling is shown to contribute significantly to the THz response.
\end{abstract}


\maketitle


Terahertz (THz) field induced transport effects can strongly influence the behavior of fast semiconductor components operating under extreme conditions \cite{Ganichev:06,Carr:02}.
The internal dynamics of the charge carriers in such devices is, among other things, determined by the band structure of the solid.
Consequently, a complete understanding of the underlying physical processes requires a detailed knowledge of the band structure.
In recent years, the development of extremely strong pulsed THz sources and ultrafast coherent detection methods \cite{Bartel:05,Gaal:06,Shen:07,Yeh:07,Blanchard:07,Dai:09,Hebling:08,Leinss:08} has made it possible 
to continuously alter the lattice momenta of the electrons in the solid and, thus, to explore their properties within the entire Brillouin zone.
This can potentially yield new tools for characterizing the band structure and analyzing transport kinetics.

Using various techniques of time-resolved THz nonlinear spectroscopy 
\cite{Gaal:06,Shen:07,Leinss:08,Shen:08,Danielson:07,Gaal:08,Wen:08,Razzari:09,Su:09,Hoffmann:09,Hebling:10,Feng:09,
Wright:09,Kuehn:10,Hoffmann:10,Kuehn:10_1,Liu:10,Hirori:10,Ogawa:10}, 
effects like THz-induced intervalley scattering \cite{Razzari:09,Su:09} and ballistic transport of electrons across half of the Brillouin zone \cite{Kuehn:10} have been investigated in semiconductors.
Currently, also THz-pump/THz-probe (TPTP) techniques seem very attractive in probing nonlinear carrier dynamics induced by intense few-cycle THz pulses in semiconductors. 
For instance, TPTP has been applied to induce impact ionization in InSb \cite{Wen:08,Hoffmann:09}
as well as intervalley scattering in doped GaAs, Si, and Ge \cite{Hebling:10}.

\begin{figure}[b]
\includegraphics[width=0.43\textwidth]{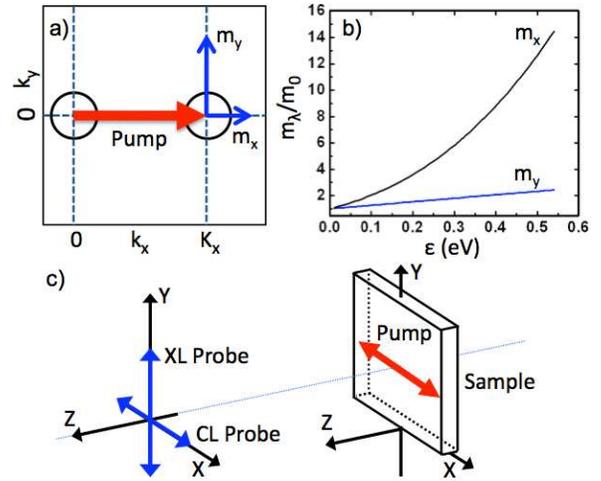} 
\caption{(color online) 
(a) The electric field of the THz pump pulse drives the electronic distribution (black circles) high in the band to an average momentum of $\k=(K_x,0)$, where the THz probe pulse senses the effective mass $m_\mr{x}$ or $m_\mr{y}$ in the $x$- or $y$-direction, respectively.
(b) Effective mass normalized to the effective mass at the bottom of the band as function of the electron energy $\veps^{}_{K_x}$ for co- (black line) and cross-linear (blue line) configurations.
The effective mass is computed from Eqs.~(\ref{eq:bands}) and (\ref{eq:masses}) using a nonparabolicity factor of $\alpha=1.33\,\mr{eV}^{-1}$.
(c) Schematic of a polarization dependent THz-pump/THz-probe (TPTP) experiment. 
For the co-linear (CL) configuration, both THz pulses are polarized in the $x$-direction whereas the probe polarization is along the $y$-axis for cross-linear (XL) TPTP.
}
\label{fig:schematic}
\end{figure}

For III-V semiconductor compounds, the conduction band energy $\veps_\k$ is parabolic and symmetric only for low electronic momenta $\hbar\k$ implying the same effective electron mass in all directions, as illustrated in Fig.~\ref{fig:schematic}(a) at $\k=(0,0)$. 
As the electron distribution moves towards higher momenta at $\k=(K_x,0)$, however, $\veps_\k$ becomes strongly nonparabolic \cite{Lundstrom:00} producing strongly directional, i.e.\ anisotropic, effective masses, as shown in Fig.~\ref{fig:schematic}(b) along the $x$- (black line) and $y$- (blue line) directions for a typical isotropic, nonparabolic band as a function of the corresponding electron energy.
In other words, due to the nonparabolicity of $\veps_\k$, the two components of the effective mass are equivalent only at the bottom of the band but they strongly deviate from each other at higher carrier momenta.

Traditionally, the effective mass is measured using magnetic cyclotron resonance (CR) \cite{Palik:61,Nicholas:85}. 
In particular, nonparabolicity was first observed in bulk InSb using CR \cite{Palik:61}, and the same technique was used to study nonparabolicity in bulk In$_{0.53}$Ga$_{0.47}$As \cite{Nicholas:85}, showing an increase in the electron effective mass as a function of CR energy (up to 30 meV at 4.2 K). 
However, CR probes the average effective mass (often called cyclotron mass) for a given orbit in $k$-space and for a certain level of energy, but is not capable of probing the above mentioned anisotropy of the effective mass at a specific point in $k$-space.

In this Letter, we present a new technique to directly monitor the anisotropic effective mass of the electrons.
We implement a polarization dependent TPTP experiment as shown schematically in Fig.~\ref{fig:schematic}(c).
Here, a strong THz pump pulse accelerates the electrons in the $x$-direction which are then probed by another weaker THz pulse polarized either in the $x$- (co-linear) or $y$- (cross-linear) direction.
We show that the anisotropy in the electron masses yields distinctly different THz responses for the co- and cross-linear configurations because the measured THz probe signal is proportional to the inverse effective mass in the probed direction.
While the nonlinear THz response in doped semiconductors arising from band nonparabolicities has already been reported \cite{Mayer:86,Markelz:96}, our experimental results demonstrate the anisotropic nature of band nonparabolicity. 

\begin{figure}
\includegraphics[width=0.42\textwidth]{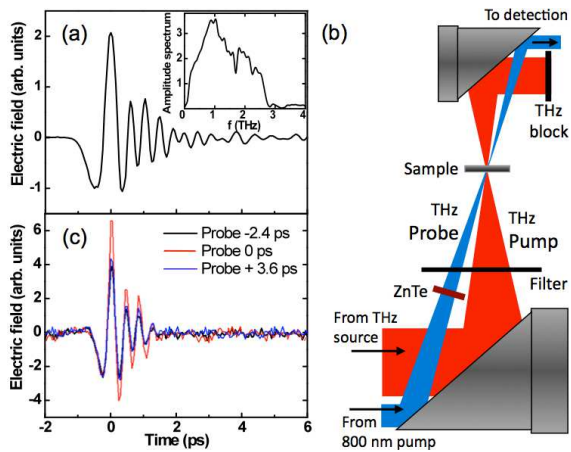}
\caption{(color online) Experimental setup. 
(a) Electric field profile of the THz pump beam emitted by the ZnTe optical rectification source.
Inset: amplitude spectrum of the THz pulse. 
(b) Schematic of the experimental setup.
(c) Electric field profile of the transmitted THz probe beam at various delay times between the main positive peaks of the THz pump and THz probe pulses.
}
\label{fig:exp1}
\end{figure}

In our experiments, a large aperture ZnTe optical rectification source \cite{Blanchard:07} was used to generate high-power, few-cycle, THz pump pulses with energies of $0.8\,\mr{\mu J}$ and $0.1 - 3 \,\mr{THz}$ bandwidth. 
Figure~\ref{fig:exp1}(a) shows an example of the temporal profile of the THz pulses produced by the ZnTe source, and the inset shows the corresponding amplitude spectrum. 
The TPTP setup is shown schematically in Fig.~\ref{fig:exp1}(b). 
The sample used here was a 500 nm-thick $n$-type In$_{0.53}$Ga$_{0.47}$As epilayer (carrier concentration of approximately $2\times10^{18}$ cm$^{-3}$) on a 0.5-mm-thick semi-insulating InP substrate. 
A $10\times10\times0.5$\,mm$^3$ ZnTe crystal placed just after the first off-axis parabolic mirror and just before the sample was used to generate a THz probe beam that overlaps the THz pump beam at the focus on the sample. 
An additional ZnTe crystal 0.5 mm thick was used to detect the THz probe pulses transmitted through the sample by free-space electro-optic sampling. 
The spot size diameters on the sample for the THz pump beam and THz probe beam were 1.6 mm and 2.5 mm, respectively (the probe beam path has a larger f-number than the THz pump beam).
Both the ZnTe source crystal and the ZnTe detection crystal for the THz probe beam could be rotated to produce (and detect) probe polarization states either parallel or perpendicular to the THz pump beam. 
At the sample position, the THz pump and THz probe peak electric fields were estimated to be about 200 kV/cm and 2 kV/cm, respectively. 
The THz probe itself is in the low-field regime and therefore does not induce any nonlinear response in the sample \cite{Razzari:09,Su:09}.
We note that the non-collinear geometry of the TPTP experiment allowed the THz pump and probe beams transmitted through the sample to be geometrically separated. 
Cross-talk between the two THz beams was therefore avoided by simply placing a metallic beam block in the path of the transmitted THz pump beam after the second off-axis parabolic. 
In addition, lock-in detection of the transmitted THz probe pulse amplitude was synchronized to an optical chopper inserted in the THz probe source beam. 
As shown in Fig.~\ref{fig:exp1}(c), the amplitude of the transmitted THz probe waveform increases when it overlaps with the THz pump pulse at zero relative time delay, while the phase is relatively unaffected. 
This allows the transmission of the main positive peak of the THz probe pulse to be monitored as function of pump-probe delay time. 
We also note that all the experiments were performed under a dry nitrogen purge at room temperature.

\begin{figure}
\includegraphics[width=0.41\textwidth]{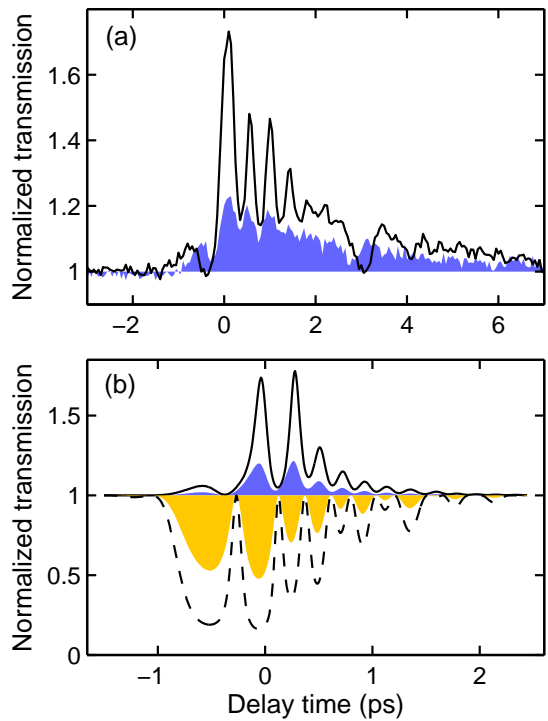}
\caption{(color online) Measured (a) and calculated (b) normalized peak transmission of the THz probe pulse as function of the pump-probe delay.
The solid lines and the blue shaded areas show the results for co- and cross-linear TPTP configurations, respectively.
The dashed line and the yellow shaded area in (b) represent corresponding calculations omitting the radiative backcoupling effects.
}
\label{fig:exp-th}
\end{figure}

Figure~\ref{fig:exp-th}(a) shows the normalized transmission of the main peak of the THz probe pulse as a function of pump-probe delay time, which is a common method for probing ultrafast carrier dynamics in semiconductors in optical-pump/THz-probe experiments \cite{Su:09,Ralph:96,Cooke:04}. 
The presence of the THz pump pulse results in an increase in transmission of the peak electric field of the THz probe pulse. 
The blue shaded area in Fig.~\ref{fig:exp-th}(a) shows the transmission change for cross-linear polarization of pump and probe beams, while the black line shows the same measurement performed for the co-linear polarization configuration. 
In the latter case, a fast, large amplitude oscillation is observed on top of a slower component similar to that shown as the area plot. 
As previously mentioned, we have excluded the possibility that these fast oscillations are due to cross-talk between the pump and the probe at the detection level. 
In particular, when the probe beam is blocked, no residual signal from the pump beam is detected; both beams have to be present inside the sample in order to observe the large amplitude oscillations for co-linear polarization shown in Fig.~\ref{fig:exp-th}(a). 
Moreover, moving the sample to an off-focus position (or completely removing it from the THz beam) eliminated any modulation signal of the THz probe transmission. 
Finally, simple interference cannot justify the observed anisotropy, due to the negligible contribution of the probe signal to the overall electric field amplitude. 
As discussed earlier for Fig.~\ref{fig:exp1}(c), the lack of any significant phase shift in the transmitted THz waveform implies that the observed signal is due to modulation of the peak amplitude of the THz probe pulse. 
To further corroborate this conclusion, no signal was observed as a function of pump-probe delay time when the THz detection point was set to a zero-crossing of the THz probe waveform.  

We next apply a microscopic theory to rigorously reveal the microscopic origin of the detected anisotropy.
The propagation of a THz field $\mb{E}(z,t)$ follows from the wave equation
\begin{equation}
\left[
\frac{\partial^2}{\partial z^2} - \frac{n^2_\mr{b}}{c^2}\frac{\partial^2}{\partial t^2}
\right] \mb{E}(z,t) = \mu_0\,\delta(z)\,\frac{\partial}{\partial t}\,\mb{J}(t),
\label{eq:wave-equation}
\end{equation}
where $n_\mr{b}$ is the background refractive index, $c$ is the speed of light, $\mu_0$ is the vacuum permeability, and $\mb{J}(t)$ is the excited current density. 
Since the thickness of the sample is much smaller than the THz field's wave length, the $z$-dependence of $\mb{J}$ is very accurately described via a $\delta$-function.
The temporal evolution of $\mb{J}$ follows then from
$\mb{J}(t) = -\frac{\el}{\hbar {\cal V}}\sum_\k (\nabla_\k\veps_\k) \,f_\k$ 
where $f_\k$ defines the microscopic electron distribution and ${\cal V}$ is the quantization volume of the semiconductor.
For not too large $k$-values, the conduction band energy dispersion is well-described by the relation \cite{Lundstrom:00}
\begin{equation}
\veps_\k\,(1+\alpha\,\veps_\k) = \frac{\hbar^2\, \k^2}{2 m^\ast},
\label{eq:bands}
\end{equation}
where $\alpha$ is the nonparabolicity factor of the band and $m^\ast$ is the effective mass at the bottom of the band. 
The $k$-dependent effective mass, i.e.
\begin{equation}
m^{-1}_i(\k) \equiv \frac{1}{\hbar^2} \frac{\mr{d}^2 \veps_\k}{\mr{d} k_i^2},
\label{eq:masses}
\end{equation}
in Fig.~\ref{fig:schematic}(b) is computed using Eq.~(\ref{eq:bands}) with the standard InGaAs value 
$\alpha=1.33\,\mr{eV}^{-1}$ \cite{Ahmed:85}.
To compute the dynamics of the electron distribution, we use an equation-of-motion approach \cite{Haug:09,Kira:06} and finally obtain
\begin{equation}
\hbar\frac{\partial }{\partial t}f_\k = \el\mb{E}(0,t)\cdot\nabla_\k f_\k
+ \left.\hbar\frac{\partial}{\partial t}f_\k\right|_\mr{scatt}.
\label{eq:eom_f}
\end{equation}
The first term on the right hand side of Eq.~(\ref{eq:eom_f}) yields the acceleration of the electrons due to the THz field.
This produces a time-dependent displacement of the electronic distribution in $\k$-space according to 
$f_\k(t) = f_{\tilde{\mb{k}}(t)}$ with $\frac{\mr{d}}{\mr{d} t}\tilde{\mb{k}}(t)=-\frac{\el}{\hbar}\mb{E}$,
according to the classical acceleration theorem \cite{Kittel:63}. 
The second term of Eq.~(\ref{eq:eom_f}) includes Coulomb and phonon scattering of the accelerated electrons.
This term leads effectively to a relaxation of the carriers and, thus, to a damping of the current density.
As we will discuss below, however, the experimental results are almost not affected by these relaxation processes. 
Consequently, we can omit the corresponding term in the equation of motion. 
This reduces the numerical complexity and allows us to perform the calculations for a true three-dimensional system.

Equation~(\ref{eq:wave-equation}) shows that the THz field entering the equation for $f_\k$ [Eq.~(\ref{eq:eom_f})] itself depends on the electron dynamics via the current density.
Consequently, Eqs.~(\ref{eq:wave-equation}) and (\ref{eq:eom_f}) must be solved self-consistently in order to account for the backcoupling of the induced fields to the carrier dynamics.
Physically, this backcoupling leads to a strong reflection of the THz field from the sample and to a radiative damping \cite{Kira:06} of the induced current density.
For the conditions used in the experiment, we find effective radiative damping/relaxation times as short as 50\,fs implying a dominance of the radiative decay. 
Consequently, the results are only weakly affected by other scattering mechanisms.

We compute the transmission of a weak THz probe pulse in the presence of a strong pump field by numerically solving Eqs.~(\ref{eq:wave-equation}) and (\ref{eq:eom_f}) for both the co- and cross-linear TPTP configurations.
We start from a Fermi-Dirac distribution for $f_\k$ at 300\,K and an electron density of 
$2\times10^{18}\,\mr{cm}^{-3}$.
Similar to the experiment, we record the peak transmission of the probe pulse for various pump-probe delay times.
The results are presented in Fig.~\ref{fig:exp-th}(b) and are in good qualitative agreement with the measured data [Fig.~\ref{fig:exp-th}(a)].
We find that $\mb{J}(t)$ produces a high reflection of the THz probe pulse with only about 5\% transmission in the absence of the pump pulse.
At the same time, the magnitude of the current induced by the probe pulse is directly proportional to the inverse effective mass. 
Therefore, the acceleration of the electrons into the nonparabolic regions can significantly modify $\mb{J}(t)$ and, thus, also the probe transmission detected. 
In particular, the solid line in Fig.~\ref{fig:exp-th}(b) shows that the transmission is strongly enhanced when pump and probe are co-linearly polarized due to the increased effective mass.
The cross-linear case (blue shaded area) produces a much smaller modification of the transmission since the effective mass is changed only moderately [see Fig.~\ref{fig:schematic}(b)].
In both cases, the fast oscillations directly reflect the transient pump-induced changes of the effective mass of the electrons.
The theory analysis unambiguously assigns the difference of co- and cross-linear excitations to the mass anisotropy.  The slow decay --- only detected in the experimental signals --- results from intervalley scattering \cite{Razzari:09,Su:09} which is not included in our model because it does not affect the detected anisotropy.

The importance of the radiative backcoupling effects can be highlighted with an artificial calculation where we neglected these effects.
Here, we simply replaced the full THz field in Eq.~(\ref{eq:eom_f}) by the external incident field.
The corresponding results are shown in Fig.~\ref{fig:exp-th}(b) as the dashed line (co-linear) and the yellow shaded area (cross linear).
Evidently, such a simplification produces completely opposite behavior than observed in the experiment.
Although the basic findings of the experiment can easily be understood intuitively by the change of the effective masses, a complete understanding requires the inclusion of the complex self-consistent backcoupling effects.

In conclusion, we have used polarization-dependent THz-pump/THz-probe techniques to study the nonlinear THz response of electrons in $n$-doped InGaAs. 
As is unambiguously confirmed by our microscopic theory, the subpicosecond time resolution of our technique, coupled with the control of probe polarization, reveals the anisotropy in effective mass of hot electrons caused by the nonparabolicity of the conduction band.
This new tool may open the way to directly mapping energy bands in semiconductors. 
Moreover, we have shown that the inclusion of backcoupling effects is crucial to understand the results.

\begin{acknowledgments}
We wish to acknowledge financial support from the Natural Sciences and Engineering Research Council of Canada (NSERC), NSERC Strategic Projects, and INRS\@. 
F.~Blanchard wishes to acknowledge le Fonds Qu\'eb\'ecois de la Recherche sur la Nature et les Technologies (FQRNT, \#138131). 
We are also thankful to Antoine Laramée and François Poitras for their technical assistance with the ALLS laser source.
\end{acknowledgments}


%

\end{document}